\documentclass[]{interact}

\usepackage{caption}
\usepackage{subcaption}
\usepackage{booktabs}
\usepackage{amsmath, amsthm, amsbsy, amsfonts, mathtools}
 \usepackage{array, multirow}
\usepackage{graphicx}
\usepackage[table]{xcolor}

\usepackage{bbm}
\usepackage{cancel}

\usepackage{amsmath}

\begin{document}
\title{A Functional Data Analysis Approach to Evolution Outlier Mining for Grouped Smart Meters}

\author{A.~Elías,
        J.~M.~Morales,%~\IEEEmembership{Fellow,~OSA,}
        ~and~S.~Pineda,%~\IEEEmembership{Life~Fellow,~IEEE}% <-this % stops a space
\thanks{A. Elías and J. M. Morales are with Dep. of Applied Mathematics, Univ. of Malaga, Spain. E-mails: aelias@uma.com; juan.morales@uma.es. S. Pineda is with the Dep. of Electrical Engineering, Univ. of Malaga, Spain. E-mail: spinedamorente@gmail.com.}% <-this % stops a space
\thanks{This work was supported in part by the Spanish Ministry of Science and Innovation through project PID2020-115460GB-I00, and in part by the Andalusian Regional Government through project P20-00153, and in part by the Research Program for Young Talented Reseachers of the University of Málaga under Project B1-2020-15.
This project has also received funding from the European Social Fund and the European Research Council (ERC) under the European Union’s Horizon 2020 research and innovation programme (grant agreement No 755705). The authors thankfully acknowledge the computer resources, technical expertise, and assistance provided by the SCBI (Supercomputing and Bioinformatics) center of the University of Málaga}% <-this % stops a space
\thanks{}
}

\maketitle

\begin{abstract}
Smart metering infrastructures collect data almost continuously in the form of fine-grained long time series. 
These massive data series often have common daily patterns that are repeated between similar days or seasons and shared among \emph{grouped meters}. 
Within this context, we propose an unsupervised method to highlight individuals with abnormal daily dependency patterns, which we term \emph{evolution outliers}.
To this end, we approach the problem from the standpoint of Functional Data Analysis (FDA)
and we use the concept of \emph{functional depth} to exploit the dynamic group structure and isolate individual meters with a different evolution.
The performance of the proposal is first evaluated empirically through a simulation exercise under different evolution scenarios.
Subsequently, the importance and need for an evolution outlier detection method is shown by using actual smart-metering data corresponding to photo-voltaic energy generation and circuit voltage records. Here, our proposal detects outliers that might go unnoticed by other approaches of the literature that have demonstrated to be effective capturing magnitude and shape abnormalities.
\end{abstract}

% Note that keywords are not normally used for peer review papers.

% \fi
\maketitle

\section{Introduction}

%\IEEEpubidadjcol

Smart metering infrastructures are spreading and with them the ability to improve the quality, efficiency, and sustainability of electricity systems. Nowadays, numerous features such as energy consumption, household  circuit voltage, and photo-voltaic energy generation are available for long time periods, at a very high-frequency rate. Furthermore, these features are  contemporaneously collected for a multitude of grouped meters. For example, residential smart meters record data from different households in a given neighborhood or city \cite{pecanStreet}. Another example is a solar energy farm collecting power generation data at the inverter level, providing as many time series as inverters \cite{solarFarm}.
This data ecosystem provides not only big data but complex data structures that requires of new advance methodologies \cite{bigdata2014, sangalli2018, sangalli2020}.

Within smart metering data analysis, outlier detection has become a topic of high interest  \cite{IEEEsun2018}. Additionally to its application to data quality \cite{AngeloIEEE2011, neuro2016, jindalIEEE2016, energyReports2020}, outlier detection methods stand out due to its capability to monitor abnormalities and discover hidden patterns.
Methodologies with this aim have been termed as \emph{outlier mining} methods \cite{IEEEsun2018} and have been useful to reveal consumer behavior, capture energy theft, find system vulnerabilities and failures, and improve service quality \cite{AngeloIEEE2011, energyReports2020, TanasaIEEE2004,vallakati2015, robinson2019, proceedingsIEEEwang2019}.

%In this article, we propose a unsupervised and non-parametric outlier detection method to monitor and mine abnormal behaviours for smart meters with a focus on interpretability.
%Particularly, the proposal is intended to set of meters smart meters records are data For example, residential smart meters record data from different households in a given neighborhood or city \cite{pecanStreet}. 
%Another example is a solar energy farm collecting power generation data at the inverter level \cite{solarFarm}. 

%\boldblue{Many methodologies to this end}
The literature of outlier detection is vast and surveys on methodologies and applications have classified the literature by groups of data analytic methodologies \cite{IEEEsun2018, reviewOutliersTS2021, appliedEnergy2021}. 
Recently, the authors in \cite{appliedEnergy2021} have provided an extensive taxonomy of the existing algorithms based on the different modules and parameters adopted, such as machine learning algorithms, feature extraction approaches, anomaly detection levels, computing platforms, and application scenarios.
From the point of view of time series analysis, in \cite{reviewOutliersTS2021} the authors propose a classification by the type of input data (univariate or multivariate time series), outlier type (point, subsequence, or time series), and nature of the method (univariate or multivariate).
Specific to the context of smart metering, the authors in \cite{IEEEsun2018} classify the available approaches into distance-based methods, density-based methods, Support Vector Machines (SVM) methods and hybrid methods.

\begin{figure}
    \centering
    \includegraphics[width=4in]{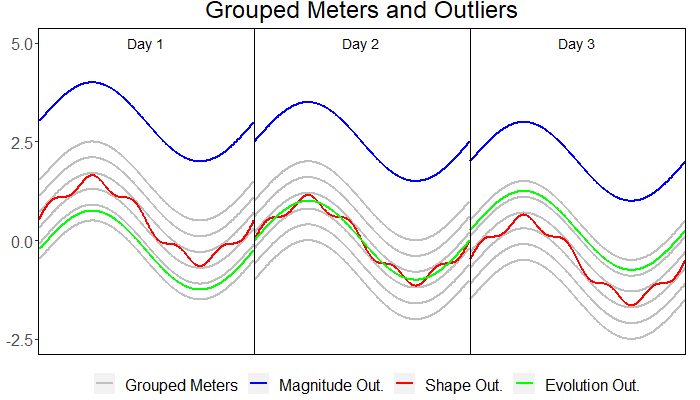}
    \caption{Taxonomy of outliers.}
    \label{fig:illustration_outliers}
\end{figure}

%\boldblue{These fail to detect important abnormalities}
The above methodologies have been successfully used in many applications to detect abnormal phenomena of importance for smart meters problems. 
In fact, methods based on L-$p$ distances are remarkable tools to identify \textit{magnitude outliers}.
For instance, the blue plot in Figure~\ref{fig:illustration_outliers} represents an outlier of this kind since it is always far away from the majority and, therefore, it features an abnormally large $\text{L-}p$ distance with respect to the other curves.
This figure also includes other two types of outliers. The red profile represents a \textit{shape outlier}, since it is not far away from the majority, but it exhibits more wiggles than the other curves. 
The green one is what we have termed an \textit{evolution outlier}, because it has similar magnitude and shape as the majority, but its day-to-day variation is quite particular. 
Note that the green plot increases from day to day, while the other curves decrease. 
%As a result, the green plot climbs positions with respect to its peers day after day.
As a result, the green plot evolves abnormally with respect to its peers day after day.
One important drawback of methods built on L-$p$ distances is that they might fail to detect both shape \cite{marron1995} outliers and individuals evolving in time differently to the rest of the members of the group  (i.e., what we have called evolution outliers) \cite{ranaAneiros2015}.

% \textcolor{red}{The above methodologies have been successfully used in many applications at detecting abnormal phenomenons of importance for smart meters analysis. These existing methods compute discrepancies among data profiles using L-$p$ distances and, consequently, are able to correctly identify \textit{magnitude outliers}. For instance, the blue plot in Figure \ref{fig:illustration_outliers} represents a magnitude outlier since it is always far away from the majority and, therefore, it presents an abnormally large L-$p$ distance with respect the other curves. This Figure also includes other two types of outliers. The red plot represents a \textit{shape outlier} since it is not far away from the majority but it presents more wiggles than the other curves. The green plot is defined as an \textit{evolution outlier} since it has similar magnitude and shape as the majority, but its day-to-day variation is quite particular. Note that the blue plot increases from day to day, while most curves decrease from day to day. One important drawback of distance-based methods is that they fail to detect both shape and evolution outliers \cite{marron1995}.}

%\boldblue{FDA does much better job}
To overcome the difficulties in detecting shape outliers (like the red plot in Fig.~\ref{fig:illustration_outliers}), the technical literature includes outlier detection methods based on Functional Data Analysis (FDA) \cite{ramsay2005functional, ferraty2006, sangalli2020}.
This branch of Statistics puts the focus on the morphological aspects of the observed curves, such as magnitude, shape and derivatives, simplifying the task of outlier detection and the interpretation of the outcomes. 
In this context, the authors of \cite{Outliergram} propose the \emph{outliergram}, which exploits the parabolic relationship between a functional depth measure and the Modified Epigraph Index \cite{pintadoromo2011} to identify shape outliers.
%Towards magnitude outlier detection, reference \cite{SunGenton11} uses the concept of functional depth to detect magnitude outliers through .

% \textcolor{red}{To overcome the shortcomings of distance-based methods, the technical literature also includes outlier detection methods that use Functional Data Analysis (FDA) \cite{ramsay2005functional, ferraty2006}. This branch of Statistics puts the focus on the morphological aspects of observed curves, such as magnitude, shape and derivatives, simplifying the task of outlier detection and the interpretation of the outcomes. For instance, reference \cite{SunGenton11} proposes a method that uses the concept of functional depth to detect magnitude outliers through the functional boxplot. In the same line, the authors of \cite{Outliergram} propose the outliergram that exploits the parabolic relationship between a functional depth measures and the Modified Epigraph Index (MEI) to identify shape outliers, such as the red plot in Figure \ref{fig:illustration_outliers}.}

%\boldblue{FDA is currently lacking skills to identify outlying evolution of the curves}
%Yet, neither the outliergram nor any of the available methods and approaches mentioned before are unable to detect evolution outliers as the one illustrated in the green plot of Figure~\ref{fig:illustration_outliers}.
Yet, none of the available methods and approaches mentioned before are designed to detect evolution outliers (like the  green plot of Figure~\ref{fig:illustration_outliers}).
In fact, to the best of our knowledge, the plausibly daily outlying temporal evolution has been ignored so far when identifying potentially relevant outliers \cite{IEEEsun2018, IEEESGwang2019}. 
Only \cite{ranaAneiros2015} proposed a bootstrap model-based method to detect periods of abnormal behaviour, being only applicable to one single meter analysis.
In this paper, we aim at filling this gap by proposing a specific method to detect individual meters with abnormal evolution patterns from a group of meters with a common structure.
Our proposal uses the information of \emph{functional depth measures} \cite{tukey1975, gijbels2017} to exploit the group structure and isolate individual meters with a different evolution. These correspond to meters with abnormal inter-day evolution patterns or, in other words, individuals that do not follow the expected daily evolution mined from the group.

Additionally, we found that when the group of meters presents a common trend or periodical variation, the information contained in the functional depths might not be enough to capture evolution outliers with inverse behaviour.
For example, if the common trend is positive, our proposal based on functional depths could still miss evolution outliers with a negative trend. 
To overcome this drawback, we propose an enrichment of functional depth measures by incorporating the information of the Modified Epigraph Index. 

% \textcolor{red}{Despite their advantages with respect to distance-based methods, existing FDA methods are still unable to detect evolution outliers. In fact, to the best of our knowledge, the daily outlying temporal evolution and outlying periodical variation patterns have been ignored so far to identify potentially relevant outliers \cite{IEEEsun2018, IEEESGwang2019}. In this paper, we aim at filling this gap in the technical literature by proposing a method based on \emph{functional depth measures} \cite{tukey1975, gijbels2017} to detect evolution outliers as the green plot in Figure \ref{fig:illustration_outliers}. These are robust order statistics that quantify the relative position of a function with respect to the sample. The proposal use the information of these statistics to exploit the group structure and to isolate individual meters with different \emph{evolution}. These correspond to meters with abnormal inter-day evolution patterns or, in other words, individuals that do not follow the expected daily evolution mined from the group.}

%For example, residential smart meters record data from different households in a given neighborhood or city \cite{pecanStreet}. Another example is a solar energy farm collecting power generation data at the inverter level, providing as many time series as inverters \cite{solarFarm}. 

%\boldblue{Main contributions}
Therefore, the main contributions of this work are:
 \begin{itemize}
     \item The proposal of an evolution-outlier detection method to unmask meters with abnormal evolution patterns based on functional depth measures. 
     \item The proposal of a depth-measure transformation that distinguishes between increments and decrements produced by trends, and peaks and valleys produced by different seasons, so that these key features of the original smart meter time series are retained.
 \end{itemize}
 
The rest of the paper is organized as follows. 
Section~\ref{sec:methodology} introduces the notation and  definitions required to describe the methodology. 
Our evolution-outlier detection method is presented in Section~\ref{sec:toolbox}. 
%Then, in Section~\ref{sec:simulresults}, we discuss results from a simulation study where we compare our methodology with general-purpose outlier detection methods of the literature under different scenarios. 
Then, in Section~\ref{sec:simulresults}, we discuss results from a simulation study where we show the empirical superiority of our proposal against general-purpose methods of the literature to detect outliers.
Additionally, we illustrate the use of the outlier detection methods with real case studies in Section~\ref{sec:results}.
Finally, Section~\ref{sec:conclusion} draws some conclusions and outlines some avenues for further research.

\section{Theoretical framework and definitions}
\label{sec:methodology}

\subsection{From smart meter data to functional data}
Let $\{\Gamma(u)$, $u \in [1, p \times T]\}$ be one meter's feature that is recorded at $p \times T$ points during $T$ windows (e.g. days) with a (daily) seasonality of length $p$. 
Then we consider each complete window record as a discrete realization of the functional process, 
\begin{align}
\label{eq:FTS}
%y_t(x) = \{X(u), \quad u \in (p(t - 1), pt) \},
y^t(x) = \{\Gamma(u), \quad u = x + p(t-1)\}, 
\end{align}
\[t \in 1, \dots, T, \quad 1 \leq x \leq p,\]
where $t$ represents the index of windows (days) and $x \in [1, p]$ is the functions' domain of definition. In the context of smart meters, the domain is usually a range covering the twenty four hours of a day, typically from midnight to midnight. The result is a series of
daily curves indexed in time by $t = 1, \dots, T$, that is, a \emph{Functional Time Series} (FTS) \cite{hormann2012}. In what follows, we stick to this setup.
Importantly, modeling each sample as a curve allows us to take advantage of the functional nature of the data. This means, for example, that we can work with the first derivatives of the curves, which, as we illustrate in the case study of photo-voltaic energy generation of Section~\ref{sec:results}, can provide valuable information for outlier detection purposes.

Many meters provide many FTS, such as the one introduced in Equation~\eqref{eq:FTS}. This data context can be framed into what is termed in the literature of FDA as a High Dimensional Functional Time Series \cite{hanlinGao2017, hanlinGao2019}. 
Let $i = 1, \dots, N$ be the index of the meters. Then, a sample of high dimensional functions takes the following form:
\[
\textbf{y}(x) = \begin{bmatrix}
y_1^1(x) & y_1^2(x) & \ldots  & y_1^T(x)\\
y_2^1(x) & y_2^2(x) & \ldots & y_2^T(x)\\
\vdots         &     \vdots     & \ddots & \vdots \\
y_N^1(x) & y_N^2(x)  & \ldots  & y_N^T(x)
\end{bmatrix}.
\]

\begin{figure}[!t]
	\centering
	\includegraphics[width=0.75\textwidth]{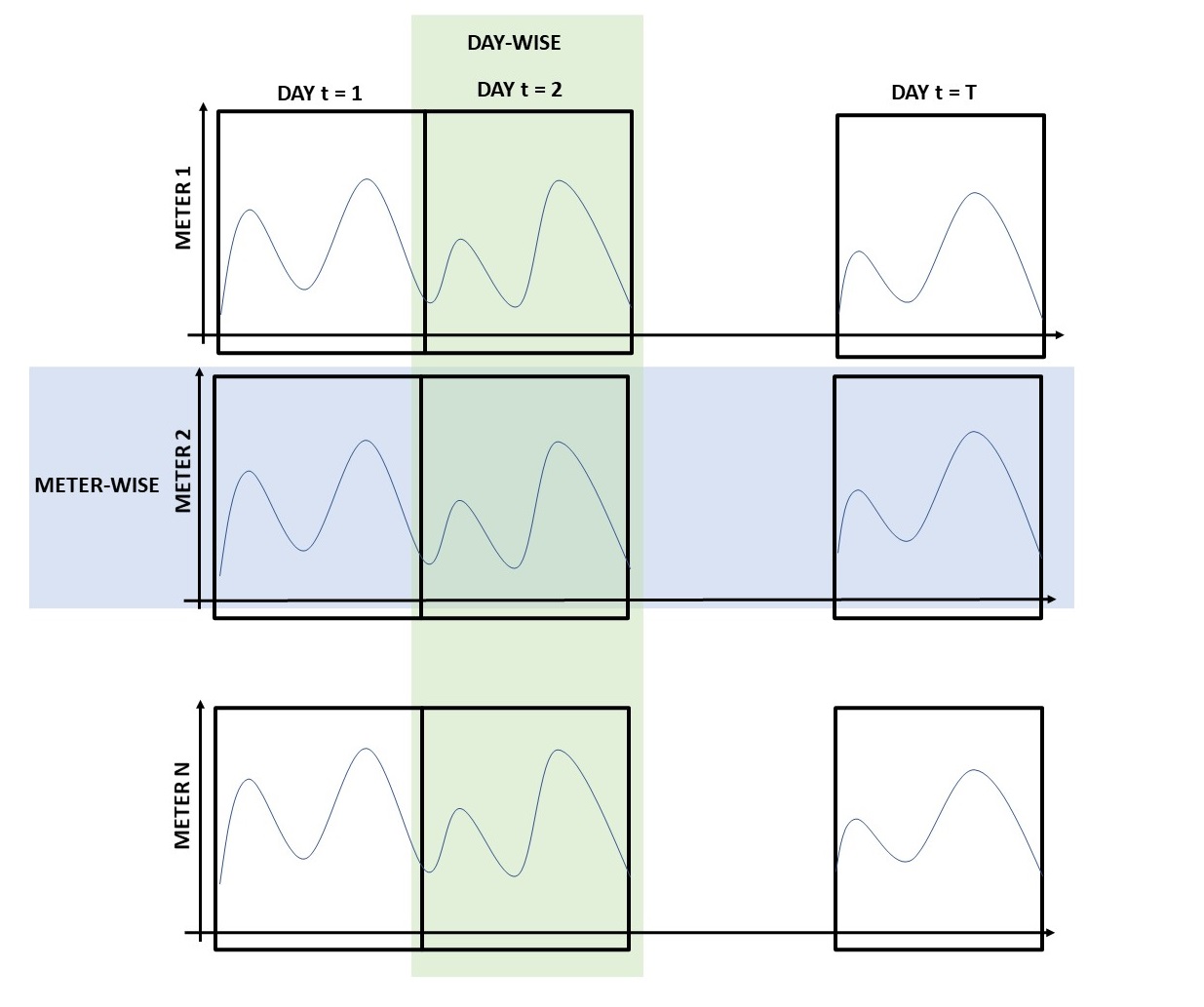}
	\caption{Context---Data from multiple smart meters.}
	\label{fig:smartmeters_illustration}
\end{figure}

Figure~\ref{fig:smartmeters_illustration} visualizes in a nutshell the information contained in the mathematical object $\textbf{y}(x)$. 
Each row represents the information of one individual meter $i$ (meter-wise), denoted by $y_i^{\cdot}(x)$. This is the FTS of $y_i^1(x), \dots, y_i^T(x)$ daily functions for one meter as in Equation~\eqref{eq:FTS}. 
On the other hand, each column represents the information of a single day $t$ (day-wise), denoted by $y^{t}_{\cdot}(x)$. This is a sample of $N$ daily functions $y^t_1(x), \dots, y^t_N(x)$ where each one represents a meter for the same day.
In the following, to ease notation, we denote by $y^1(x), \dots, y^T(x)$ the sample of $T$ functions for a given meter (meter-wise) and by $y_1(x), \dots, y_N(x)$ the sample of $N$ functions for a given day (day-wise).

% \subsection{From discrete data to smooth functions and derivatives}
% \label{subsec:derivatives}
% Meters record data discretely and, therefore, we have discrete versions of the functions. 
% For this reason, the first step in FDA is to represent each observation by the following  linear expansion
% \[y_t(x) \approx \sum_{k=1}^K c_k \phi_k (x),\]
% being $c_k$ constants and $\phi_k(x)$ a set of given basis functions. 
% The coefficients of the expansion $c_k$ can be estimated by minimizing the least squares criterion and the set of basis functions must be selected by the user, being common choices the Fourier basis or polynomial splines \cite{ramsay2005functional, ferraty2006}. 
% %\boldblue{This is needed to estimate the derivatives.}
% The step above is not only useful to get smooth functions and remove noise but to get derivative estimates. If the $ith$-derivative of the basis functions is defined, one can estimate the derivatives of the functions as follows
% \[\partial^i_x y_t (x) \approx \sum_{k=1}^K c_k \partial_x^i \phi_k (x), \]
% being $\partial_x^i $ the $i$-th differential operator with respect to $x$. As we illustrate in the case studies, the first derivatives are powerful instruments to highlight shape characteristics of the curves.

\subsection{Functional depth measures}
\label{subsec:fdepth}
The estimation of well-known statistics such as the median and quantiles are based on the ability to rank or order a data sample. 
An important property of these statistics based on rankings is that they have the ability to be insensitive to extreme observations or, in other words, they are robust. 
This fact has made them fundamental tools to construct outlier detection methods such as the classical Boxplot \cite{tukey1975}.
However, the notion of order is only unique and straightforward  in the univariate case.

To provide a notion of ordering for multivariate and high dimensional spaces, the literature has proposed the concept of \emph{depth measures} \cite{tukey1975, gijbels2017}. 
More concretely, in this article, we are interested in \emph{functional depths}, which  provide an ordering of a sample of curves and, in consequence, functional order statistics counterparts as the functional median. 

A very popular definition of functional depth is the \emph{Modified Band Depth} (MBD) \cite{pintadoromo2009}. Given a general sample of $T$ curves $Y = \{y^1(x), \dots, y^T(x)\}$, the MBD of the datum $y$ is given by
\begin{align}
    \label{eq:MBD}
&\mbox{MBD}(y; Y)  =  {T \choose 2}^{-1}\!\!\!\!\!\!\sum_{1 \leq i < j \leq T}    \frac{1}{p} \sum_{k=1}^{p} \mathbbm{1} \{ \min \{y_{i}(k), y_{j}(k) \} \leq \nonumber \\
 &\hspace{4cm}  y(k)\leq \max \{y_{i}(k), y_{j}(k) \} \}.\end{align}

\begin{figure}[!t]
	\centering
	\includegraphics[width = 0.75\textwidth]{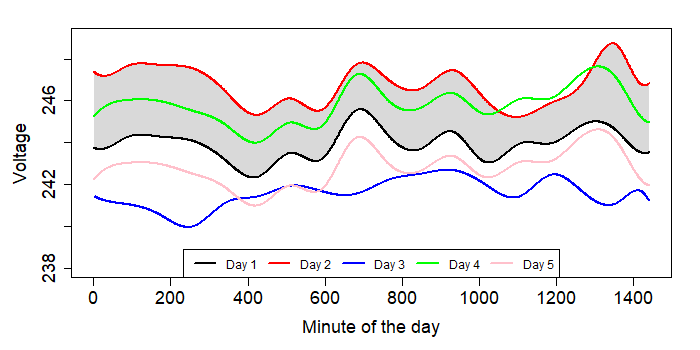}
	\caption{Synthetic voltage example for one meter for 5 days.}
	\label{fig:mbd_illustration}
\end{figure}

The equation above accounts for the mean time that the function $y$ is inside all the possible bands constructed with couples from the sample $Y$. 
This is illustrated in Figure~\ref{fig:mbd_illustration} with a synthetic row sample (see Figure~\ref{fig:smartmeters_illustration}) of smart meters data, i.e. 5 days of voltage circuit values for one household, meaning that there are ${T \choose 2} = {5 \choose 2} = 10$ possible bands. 
One of these bands is represented using the functions Day 1 and Day 2 (grey region). 
Day 4 is inside that band for a high proportion of the minutes of the day, whereas Day 3 and Day 5 are completely outside. 
Following this reasoning, notice that Day~1 is completely inside all the possible bands, thus achieving the highest depth value. 
In contrast, Day 2 and Day 3 have the two smallest depth values.

In essence, Equation~\eqref{eq:MBD} assigns a real number between $0$ and $1$ to each $y$. The highest value is the deepest function, whereas lower values correspond to observations that are outsiders with respect to the sample of functions.
Let us denote by $y^{[1]}(x),...,y^{[T]}(x)$ the ordering in terms of depths being $y^{[1]}(x)$ the deepest function and $y^{[T]}(x)$ the most outlying curve of the sample. 
The statistic $y^{[1]}(x)$ is a functional analog of the median and is a robust estimator of the center of the distribution of the functions.

Another functional statistics used in outlier detection methods \cite{Outliergram} is the Modified Epigraph Index (MEI) \cite{pintadoromo2011}.
It measures the mean proportion of curves lying above a given function $y(x)$ and is defined as
\begin{eqnarray}
\label{eq:MEI}
\mbox{MEI}(y; Y) & = & \frac{1}{T}\sum_{1 \leq i \leq T} \frac{1}{p} \sum_{k=1}^{p} \mathbbm{1} \{ y(k) \leq  y_{i}(k)\}.
\end{eqnarray}

Continuing with the example of Figure~\ref{fig:mbd_illustration}, Day 3 is the one with the highest MEI since it has  a high proportion of curves (4 out of 5) above it almost all the time. In contrast, Day 2 has the smallest MEI. 
In Section~\ref{sec:toolbox} we leverage this statistic to provide a meaningful and useful modification of functional depths to detect evolution outliers.

\section{Methodology for evolution-outlier detection}
\label{sec:toolbox}

Our idea is to exploit functional depth measures to capture the dynamic daily evolution of smart meter data. 
With this goal in mind, we use the FTS provided by each meter $i$ (meter-wise). This is $T$ daily functions for each single meter, and we compute the depth values of each of the functions in the sample of curves $Y = \{y^1(x), \dots, y^T(x) \}$.
That is, for each $t = 1, \dots, T$, we obtain $\mbox{MBD}(y^t, Y)$. Hereafter, we denote $\mbox{MBD}(y^t, Y)$, as $\mbox{MBD}(t)$, that is, the functional depth value of the day $t$ with respect to its historical records.
Then, our approach is focused on the analysis of these depths arranged as the following time series
\[\{\mbox{MBD}(t), \quad t \in (1, \dots, T)\}.\]

\begin{figure}[!t]
	\centering
	\includegraphics[width=0.75\textwidth]{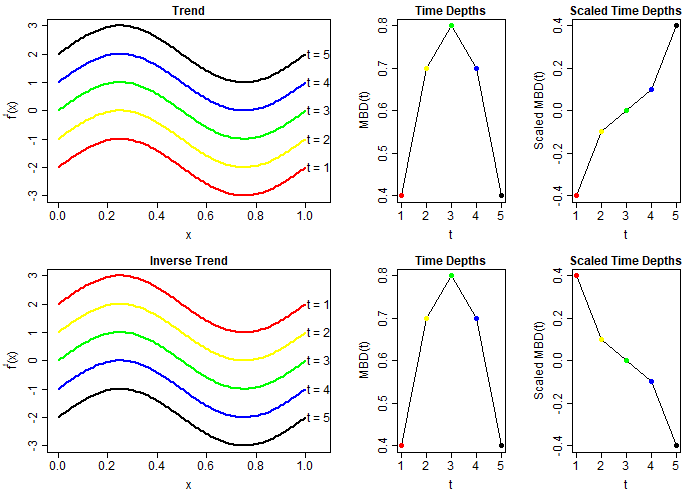}
	\caption{Time series of MBD and time series of scaled MBD for two Functional Time Series.}
	\label{fig:toy_example}
\end{figure}
To illustrate the intuition behind our proposal, see Figure~\ref{fig:toy_example}. 
The top-left panel shows a FTS that increases in magnitude from $t=1$ to $t=5$ or, in other words, it has a positive trend component. 
The deepest function is the green curve, $t=3$, and the curves with the smallest depth values are the red and the black curves, $t=1$ and $t=5$.
The top-central panel arranges depth values as a time series $\mbox{MBD}(t)$ where each point is related to one curve. 
Then, the highest depth is observed at time point $t=3$ and the two lowest are observed at $t=1$ and $t=5$.
On the other hand, the bottom-left panel illustrates the same FTS but with a decreasing trend component. That is, we invert the time order of the curves: see, for example, the position of the blue curve in the top and bottom left panels. 
The time depths $\mbox{MBD}(t)$ for this FTS with inverse trend is represented again in the central bottom panel, featuring exactly the same pattern as that of the FTS on the top panel.

%\boldblue{depths do not retain important time series features}
This toy example illustrates how depth measures are able to track the time position of each curve and how the corresponding time series of depths retain characteristics of the FTS.
However, time series of depths only account for the relative position of a function with respect to the center and do not discriminate between deviations above or below the central deepest function. 
So, two functions with the same depth value might be in opposite locations with respect to the center (see that the MBD values of the green and blue curves are the same in both the top and bottom panels).
As illustrated above, the FTS in the top panel of Figure~\ref{fig:toy_example} and the inverted FTS emphasize this problem; curves in yellow, red, black and blue have a different overall position with respect to the center, some are above and the others below. Nevertheless, the depth values are not able to capture this feature.

%\boldblue{we propose a modification of depths}
To overcome this drawback, we incorporate the information provided by the Modified Epigraph Index (MEI) to depth measures.
Using this statistic, introduced in Equation~\eqref{eq:MEI}, we define what we call the \emph{scaled depth} as
%\begin{definition}{\bf{Empirical Functional Depth} (MBD).}
\begin{align}
\widetilde{\mbox{MBD}}(y) &= sgn(\mbox{MEI}(y^{[1]})-\mbox{MEI}(y)) \label{eq:modification_MBD_sign}  \\
& \times (\mbox{MBD}(y^{[1]}) - \mbox{MBD}(y)), \label{eq:modification_MBD_center}
\end{align}
%\end{definition}
where $sgn$ stands for the sign function and $y^{[1]}$ is the functional median. 
The term~\eqref{eq:modification_MBD_sign} takes into account whether or not a function is above or below the median curve, being positive if it is above and negative otherwise. On the other hand, term~\eqref{eq:modification_MBD_center} centers the $\widetilde{\mbox{MBD}}$ in zero, being this value associated with the deepest curve. 
Now, functions below the median have a negative $\widetilde{\mbox{MBD}}$, while these are positive for functions above the median.

%\boldblue{Introduce time depth}
Analogously to depth measures, the scale depth measures provide a time series where each of the time points represents the value of a given day $t$,
\[\{\widetilde{\mbox{MBD}}(t), \quad t \in (1, \dots, T)\}.\]
%\boldblue{Intuition of scale depths with toy example}
The time series of scaled depth is defined to capture the trend and seasonal patterns of the original time series.
This is illustrated in the right-hand panels of Figure~\ref{fig:toy_example} where the scaled depth measures are plotted. Now, the information of the positive and negative trends are retained in the scaled depths. This is visually evident by an increasing and decreasing time series of scaled depths.

%\boldblue{Introduce the multivariate version of depth and transformations}
Given $N$ meters, we thus have $\mbox{MBD}_i(t)$ and $\widetilde{\mbox{MBD}}_i(t)$ for $i = 1, \dots N$. 
For simplicity, we continue the exposition for $\widetilde{\mbox{MBD}}_i(t)$ but everything can be extrapolated to $\mbox{MBD}_i(t)$.
These $N$ time series of $\widetilde{\mbox{MBD}}_i(t)$ are gathered in the following multivariate time series of scaled depths
% \[
% \widetilde{\mbox{\textbf{MBD}}}(t) = \begin{bmatrix}
% \widetilde{\mbox{MBD}}^1(1) & \widetilde{\mbox{MBD}}^2(1) & \ldots  & \widetilde{\mbox{MBD}}^I(1)\\
% \widetilde{\mbox{MBD}}^1(2) & \widetilde{\mbox{MBD}}^2(2) & \ldots & \widetilde{\mbox{MBD}}^I(2)\\
% \vdots         &     \vdots     & \ddots & \vdots \\
% \widetilde{\mbox{MBD}}^1(T) & \ldots  & \ldots  & \widetilde{\mbox{MBD}}^I(T)
% \end{bmatrix}.
% \]
\begin{equation}
\widetilde{\mbox{\textbf{MBD}}}(t) = [\widetilde{\mbox{MBD}}_1(t), \widetilde{\mbox{MBD}}_2(t), \ldots , \widetilde{\mbox{MBD}}_N(t)]. \label{eq:multivariateMBD}
\end{equation}

%\boldblue{The time series must be similar. Deviations are due to a different temporal pattern.}
Daily dependent data must result in time series in Equation~\eqref{eq:multivariateMBD} that variate in a structured way. 
Additionally, since we are focusing on meters that belong to a group, this multivariate time series must be synchronized sharing common movements among meters. 
Hence, deviations from this common evolution would determine an abnormal dependency pattern.
%\boldblue{We capture the overall dependency pattern by trimmed means}
To capture the overall time dependency pattern we compute the trimmed mean of $\widetilde{\mbox{\textbf{MBD}}}(t)$ \cite{Fraiman2001} and we use it as a robust estimate of the overall evolution. This is
\[\mu \widetilde{\mbox{MBD}}(t) = \frac{1}{\lceil N/2  \rceil} \sum_{r=1}^{\lceil N/2  \rceil} \widetilde{\mbox{MBD}}_{[r]}(t), \]
being $\widetilde{\mbox{MBD}}_{[r]}(t)$ the $r$-$th$ deepest time series of $\widetilde{\mbox{\textbf{MBD}}}(t)$.

%\boldblue{Time series far away from the overall are outliers}
Then, we use the Euclidean distance between each $\widetilde{\mbox{\textbf{MBD}}}(t)$ and the prototype $\mu \widetilde{\mbox{MBD}}(t)$ to find individuals with a time evolution that is far from the prototype evolution, i.e.,
\begin{align}
    d(\widetilde{\mbox{MBD}}_i(t), \mu \widetilde{\mbox{MBD}}(t)) = \sqrt{\sum_{t=1}^T \big(\widetilde{\mbox{MBD}}_i(t)-\mu \widetilde{\mbox{MBD}}(t)\big)^2 }. \nonumber
\end{align}
Large values of $d(\cdot, \mu \widetilde{\mbox{MBD}}(t))$ indicate that the dependency pattern is abnormal with respect to the prototype evolution. 

%\boldblue{How much is so far and how much is not so far? We need an objective threshold.}
The next step is to define a threshold or cutoff to objectively determine which is far enough away to be unmasked as an evolution outlier.
For this purpose, we study the empirical distribution of the vector of distances $\mathbf{d}$.
Intuitively, one might expect that it should be right-skewed given that we are dealing with squared values. 
However, to avoid distributional assumptions, we opt for a flexible adaptation by \cite{hubert2008} of the classical Tukey's boxplot rule. 
Precisely, we highlight a given meter $i$ as outlier if
\[ d(\widetilde{\mbox{MBD}}_i(t), \mu \widetilde{\mbox{MBD}}(t)) > Q_3(\mathbf{d}) + \gamma \times \exp^{3 MC} \times IQR(\mathbf{d}), \]
where $Q_3$ and $IQR$ are the third quantile and the interquartile range, $MC$ the medcouple statistics and $\gamma$ a parameter to tune the length of the whiskers. 
The authors of \cite{hubert2008} set it to $1.5$ to leave roughly $1\%$ of probability in both tails but, since we are only looking for right-tailed outliers, we consider $\gamma = 0.72$ to leave approximately $5\%$ only in the right tail of the distribution. 
Note that the flexibility of this rule comes from the fact that, if the distribution of $\mathbf{d}$ is symmetric ($MC = 0$), then the proposal by \cite{hubert2008} turns out to be the classical Tukey's Boxplot. Therefore, it only corrects under departures from the symmetry assumption.

In summary, we propose two methods for detecting evolution outliers. 
The first one,  which we refer to as ``TDEPTH'' for short, identifies evolution outliers based on the analysis of time series of functional depths computed as in Eq.~\eqref{eq:MBD}, while the second one, which we call ``STDEPTH,'' relies on the scaled depth that we have defined in Eqs.~\eqref{eq:modification_MBD_sign} and~\eqref{eq:modification_MBD_center}. Both of these methods can be applied to the data curves themselves and their derivatives, as we do in the numerical analysis of Section~\ref{sec:results}.

\section{Results on synthetic data sets}
\label{sec:simulresults}
\subsection{Simulation setup}
\label{subsec:setup}
The simulation setting is based on two parts. 
The first one aims to generate a set of non-atypical meters with a common evolution structure (group effect) and individual variations to each meter (meter effect).
The second part consists in generating evolution outliers by modifying the group effect and/or the meter effect components. 
In the following, we propose two different models: The first one demonstrates the ability of our proposal to detect evolution outliers by comparison against general-purpose outlier identification methods available in the technical literature. The second one shows why it is important to reformulate the standard notion of functional depth to detect some particular types of evolution outliers.

\paragraph*{Model 1:} 
This model generates a common evolution pattern for the non-outlier individuals and produces evolution outliers by adding a temporal trend.
We simulate a sample of $N$ typical meters with $T$ daily curves as follows:
\begin{align}
y_i^t(x)  = &  \sin(2 \pi x) + \tag{Group mean} \\
& \epsilon^t (x) \tag{Group effect} + \\ 
& \upsilon_i (x), \tag{Meter effect} 
\end{align}
being $\epsilon^t(x)$ and $\upsilon_i(x)$  Gaussian processes with covariance functions $Cov_{\epsilon}(x, x') =  \eta_{\epsilon} e^{-\lambda_{\epsilon} | x - x' |}$ and $Cov_{\upsilon}(x, x') =  \eta_{\upsilon} e^{-\lambda_{\upsilon} | x - x' |}$, respectively \cite{GPbook2003}. 
The parameters were set to $\eta_{\epsilon} = 0.8$, $\lambda_{\epsilon} = \lambda_{\upsilon} = 0.1$ and $\eta_{\upsilon} = 1.5$. 

To generate a meter $i$ that behaves as an evolution outlier, we select a starting time point for the trend, $t_a$, from $t \in 1,...,T-\rho$ and an end point, $t_b$, so that the trend remains during $\rho$ periods. Then, the trend is generated as a functional linear interpolation between the curve $y_i^{t_a}$ and the curve $y_i^{t_b}$. More precisely,
\begin{equation}
y_{\text{Out}1}^t(x) =
\begin{cases} 
  \frac{t_b-t}{t_b-t_a}y_i^{t_a}(x)+\frac{t-t_a}{t_b-t_a}y_i^{t_b}(x) & \text{for } t \in t_a, \dots, t_b, \\
  y_i^t(x) & \text{otherwise.} \nonumber
\end{cases}
\end{equation}
For these outliers, we fix $\eta_{\upsilon} = 0.5$. 

\paragraph*{Model 2:}
This model generates a group of meters with a common trend and includes evolution outliers by inverting the sign of the trend. 
We simulate a sample of $N$ typical meters with $T$ daily curves as follows:
\begin{align}
y_i^t(x)  = &  \sin(2 \pi x) + \tag{Group mean} \\
& \frac{T-t}{T-1}\epsilon^{1}(x)+\frac{t-1}{T-1}\epsilon^{T}(x) \tag{Group trend} + \\ 
& \upsilon_i (x), \tag{Meter effect} 
\end{align}
The parameters were set to $\eta_{\epsilon} = 0.8$ and $\lambda_{\epsilon} = 0.1$ for the group trend and $\eta_{\upsilon} = 1.5$ for the meter effect. 
The model above includes a common trend that is increasing or decreasing depending on the difference between $\epsilon^1$ and $\epsilon^T$. 

To generate an outlier with an opposite trend we revert the pattern as follows \begin{align}
y_{\text{Out}2}^t(x) = &  \sin(2 \pi x) + \tag{Group mean} \\
& \frac{t-1}{T-1}\epsilon^{1}(x)+\frac{T-t}{T-1}\epsilon^{T}(x) \tag{Outlying trend} + \\ 
& \upsilon_i (x). \tag{Meter effect} 
\end{align}

% \subsubsection{Model 3}
% This models generate a group of meters that do not share a common evolution but the outliers presents a trend.
% For each meter $i$, we simulate its $T$ curves as follows:
% \begin{align}
% y^i_t(x)  = &  \sin(2 \pi x) + \tag{Group mean} \\
% & \epsilon^i_t (x), \tag{Meter effect} 
% \end{align}
% The parameters were set to $\eta_{\epsilon} = 0.8$ and $\lambda_{\epsilon} = 0.1$ for the group trend and $\eta_{\upsilon} = 1.5$ for the meter effect. In this model, outliers are simulated as in $y^{\text{Out}2}_t(x)$.

\subsection{Benchmark methods}
Benchmark methods were chosen to cover the taxonomy of classes proposed by the survey of outliers detection methods for smart meters \cite{IEEEsun2018}. Additionally, we included large-scale unusual time series detection \cite{feature2015} and functional data analysis methods \cite{Outliergram, SunGenton11}. 
The list of final methods and their implementation are,
\begin{itemize}
    \item Distance-based methods: K-Nearest Neighbours (KNN) \cite{knnout2000} and Aggretate KNN (AKNN) \cite{aggknnout2002}. To set the parameter $K$ we try a range of values and we show the results of the best performance. Given the KNN and the AKNN scores, we determine as outlier those individuals with scores larger than $Q3 + 1.5*IQR$.
    \item Local density methods: Local Outlier Factor (LOF) \cite{lof2000}, Connectivity Based Outlier Factor (COF) \cite{cof2002} and Influenced Local Outlier Factor (INFLO) \cite{inflo2006}. Since these methods depend on the $K$-Nearest Neighbours, we follow the same procedure as with the distance-based methods to select the parameter $K$ and the outlier detection rule.
    \item One-class classification (ONESVM): We perform one-class Support Vector Machines \cite{oneclassSVM1999} classification with radial and polynomial kernels. The parameter of the radial kernel was fixed to $0.05$. We show the results of the best performing kernel. 
    \item Time series feature selection (FEA): Following \cite{feature2015}, we compute, for each meter, several time series indicators such as autocovariance features, entropy, lumpiness, flat spots, crossing points, mean, variance, maximum level shift and maximum variance shift. Then, we apply Principal Components Analysis to the data set of the features and we detect outliers in the reduced two-dimensional space produced by the first two principal components. As outlier detection methods, SVM with radial or polynomial kernel and the $\alpha-$hull method are considered. We show the results of the best performing method among SVM and $\alpha-$hull. 
    \item Dimension reduction methods (PCA): We apply Principal Components Analysis to our original data set and apply one-class SVM classification to the reduced two-dimensional space. We show the results of the best performing method among SVM and $\alpha-$hull.
    \item Functional Boxplot (FBOX): We apply the functional boxplot by \cite{SunGenton11} day-wise. A meter is highlighted as outlier if it is detected as outlier in more than $95\%$ of the days. The functional boxplot parameter is set as default.
    \item Outliergram (OUTGRAM): We apply the functional outliergram by \cite{Outliergram} day-wise and we proceed as with the FBOX.
\end{itemize}

\subsection{Performance Metrics}
To measure the performance of the outlier detection methods, we compute the True Positive Rate (TPR) and the True Negative Rate (TNR) \cite{performanceIEEE2019}, which are:
$$
\mbox{TPR} = \frac{\#\text{True Detected Outliers}}{\#\text{Outliers}},
$$
$$
\mbox{TNR} =\frac{\#\text{True Detected No Outliers}}{\#\text{No outliers}}.
$$
TPR (sensitivity) measures the fraction of anomalous events identified by a method and TNR (specifity) measures the fraction of non-anomalous events identified by the method. Therefore, the best performance would be provided by a method with $\mbox{TPR}=\mbox{TNR}=1$. In contrast, a method that is not useful to detect the outliers at all would provide $\mbox{TPR}=0$ and $\mbox{TNR}=1$.

\subsection{Results}
We generate samples with $N = 100$ and $T = 50$. Besides the $N$ meters, we add a $1\%$, $5\%$ and $10\%$ of outliers. Then, we report the mean values of TPR and TNR for $100$ replicates of this experiment. These values are collated in Tables~\ref{tab:model1} and~\ref{tab:model2} for each of the two data-generating models described in Subsection~\ref{subsec:setup}. 

Table~\ref{tab:model1} shows the results of Model~1 ($\rho=5$) and, in bold numbers, we highlight the best performing method. This is the proposed outlier detection method TDEPTH that ties with the scaled version STDEPTH, both having $\mbox{TPR} = \mbox{TNR} = 1$.
They are followed by local-density methods (LOF, COF and INFLO) but with $\mbox{TPR}$ values far away from $1$.
The results of distance-based (KNN and AKNN) and functional data methods (FBOX and OUTGRAM) show that they do not have capability of detecting these evolution outliers.
Finally, FEA and PCA seem to split the individuals into two random groups given the values of TPR and TNR around $0.5$.

The results of Model~2 are shown in Table~\ref{tab:model2}. As expected, TDEPTH does not perform well in these circumstances as motivated in Section~\ref{sec:methodology}. However, our scaled proposal, STDEPTH, achieves values for $\mbox{TPR}$  and $\mbox{TNR}$ that are again close to one, meaning that it is still able to capture this particular evolution abnormality.
For Model~2, distance-based (KNN) and local-density methods (LOF, COF and INFLO) improve their performance but they are still far from the results provided by STDEPTH.
%The improvement of these methods can be explained by the complexity of generating an outlier $y_{Out2}^t$ without falling also in the class of magnitude outliers, where these methods perform well.

In conclusion, the results in Tables~\ref{tab:model1} and \ref{tab:model2} clearly demonstrate that, among all existing outlier detection techniques in the technical literature, our methodology is the only one able to efficiently capture these evolution outliers. 
In the next Section~\ref{sec:results}, we corroborate the usefulness and importance of our approach by identifying evolution outliers that remain undetected by other methods on real data of household voltage circuit and solar energy generation.

% Additionally, Appendix~\ref{subsec:magnitude_outlier} compiles results for a curve-generating model with \emph{magnitude} outliers (see Table~\ref{tab:magnitude}). Except for OUTGRAM, TDEPTH, STDEPTH and ONESVM, all the other methods detect well this type of outlier, with an efficiency that depends on the proper selection of the meta parameters. For this additional simulation, FBOXPLOT performs the best with its default parameters, although the differences are not large. Finally, a curve-generating model with \emph{shape} outliers is shown in Appendix~\ref{subsec:shape_outlier_simul}. These are only detected by OUTGRAM (see Table~\ref{tab:shape}).

\begin{table}[]
\centering
\caption{Simulation results for Model 1.}
\label{tab:model1}
\begin{tabular}{lllllll}
\toprule
\textbf{Outliers} & \multicolumn{2}{c}{1$\%$} & \multicolumn{2}{c}{5$\%$} & \multicolumn{2}{c}{10$\%$} \\
\midrule
& \textbf{TPR} & \textbf{TNR} & \textbf{TPR} & \textbf{TNR} & \textbf{TPR} & \textbf{TNR} \\
 \midrule
 TDEPTH & \textbf{1.000} & \textbf{1.000} & \textbf{1.000} & \textbf{1.000} & \textbf{1.000} & \textbf{1.000} \\
 STDEPTH & 1.000 & 1.000 & 1.000 & 1.000 & 1.000 & 1.000 \\
 KNN & 0.030 & 0.904 & 0.010 & 0.897 & 0.008 & 0.895 \\
 AKNN & 0.000 & 0.938 & 0.010 & 0.929 & 0.004 & 0.923 \\
 LOF & 0.270 & 0.886 & 0.172 & 0.888 & 0.089 & 0.888 \\
 COF & 0.120 & 0.955 & 0.166 & 0.954 & 0.066 & 0.955 \\
 INFLO & 0.270 & 0.959 & 0.038 & 0.961 & 0.006 & 0.956 \\
 ONESVM & 0.380 & 0.676 & 0.340 & 0.699 & 0.394 & 0.672 \\
 FEA & 0.490 & 0.514 & 0.482 & 0.506 & 0.43 & 0.563 \\
 PCA & 0.490 & 0.506 & 0.482 & 0.506 & 0.43 & 0.563 \\
 FBOX & 0.000 & 0.991 & 0.000 & 0.988 & 0.000 & 0.985 \\
 OUTGRAM & 0.000 & 0.955 & 0.000 & 0.960 & 0.001 & 0.959 \\
  \midrule
\end{tabular}
\end{table}

\begin{table}[]
\centering
\caption{Simulation results for Model 2.}
\label{tab:model2}
\begin{tabular}{lllllll}
\toprule
\textbf{Outliers} & \multicolumn{2}{c}{1$\%$} & \multicolumn{2}{c}{5$\%$} & \multicolumn{2}{c}{10$\%$} \\
\midrule
& \textbf{TPR} & \textbf{TNR} & \textbf{TPR} & \textbf{TNR} & \textbf{TPR} & \textbf{TNR} \\
 \midrule
 TDEPTH & 0.07 & 0.951 & 0.054 & 0.946 & 0.039 & 0.947 \\
 STDEPTH &  \textbf{1} & \textbf{0.951} & \textbf{0.99} & \textbf{0.972} & \textbf{0.983} & \textbf{0.989}  \\
 KNN  & 0.48 & 0.902 & 0.444 & 0.920 & 0.341 & 0.918 \\
 AKNN & 0.09 & 0.932 & 0.116 & 0.944 & 0.083 & 0.942 \\
 LOF  & 0.69 & 0.882 & 0.646 & 0.908 & 0.537 & 0.908 \\
 COF  & 0.56 & 0.952 & 0.596 & 0.960 & 0.41 & 0.970 \\
 INFLO & 0.48 & 0.962 & 0.136 & 0.962 & 0.046 & 0.958 \\
 ONESVM & 0.6 & 0.677 & 0.53 & 0.711 & 0.523 & 0.718 \\
 FEA  & 0.57 & 0.507 & 0.312 & 0.764 & 0.638 & 0.517 \\
 PCA  & 0.43 & 0.574 & 0.312 & 0.764 & 0.638 & 0.517 \\
 FBOX  & 0.07 & 0.989 & 0.094 & 0.986 & 0.063 & 0.986 \\
 OUTGRAM  & 0.2 & 0.953 & 0.192 & 0.950 & 0.177 & 0.946 \\
  \midrule
\end{tabular}
\end{table}

% \begin{table}[]
% \centering
% \caption{Simulation results for Model 3.}
% \label{tab:trend}
% \begin{tabular}{lllllll}
% \toprule
% \textbf{Outliers} & \multicolumn{2}{c}{1$\%$} & \multicolumn{2}{c}{5$\%$} & \multicolumn{2}{c}{10$\%$} \\
% \midrule
% & \textbf{TPR} & \textbf{TNR} & \textbf{TPR} & \textbf{TNR} & \textbf{TPR} & \textbf{TNR} \\
%  \midrule
%  TDEPTH & 0.35 & 0.939 & & & & \\
%  STDEPTH & \textbf{0.98} & \textbf{0.949} & \textbf{} & \textbf{} & \textbf{} & \textbf{} \\
%  KNN  & 0.32 & 0.991 & & & & \\
%  AKNN & 0.32 & 0.993 & & & & \\
%  LOF  & 0.36 & 0.960 & & & & \\
%  COF  & 0.24 & 0.990 & & & & \\
%  INFLO & 0.37 & 0.953 & & & & \\
%  ONESVM & 1 & 0 & & & & \\
%  FEA  & 0.28 & 0.988 & & & & \\
%  PCA  & 0.28 & 0.988 & & & & \\
%  FBOX  & 0.07 & 1 & & & & \\
%  OUTGRAM  & 0.2 & 0.976 & &  & & \\
%   \midrule
% \end{tabular}
% \end{table}

\section{Results on real data sets}
\label{sec:results}

Next we use the advocated FDA approach to evolution outlier detection with real smart meter data. 
Additionally, to cover the complete taxonomy of outliers introduced in Figure~\ref{fig:illustration_outliers}, we also apply the functional boxplot \cite{SunGenton11} and the outliergram \cite{Outliergram}, two of the most efficient methods to detect magnitude and shape outliers.
Specifically, we use the Pecan Street data set \cite{pecanStreet} that provides access to 1-minute records of smart meters from Austin over one year. 
We use freely available data of voltage circuit (25 households) and solar energy generation (19 households\footnote{Given the metadata of the Pecan Street data set, households $8565$, $8386$, $9922$, $5746$, $7951$ and $7901$ do not have photo-voltaic energy generation.}).

In all the results, the parameters are set with their default values as explained in Section~\ref{sec:toolbox}. 
Additionally, for photo-voltaic data, we work with  the non-zero solar generation profiles, obviating night time periods. 
The FDA methods are applied to the smoothed level data and the first derivatives. To smooth the data and to estimate the derivatives, we use cubic B-splines and the number of basis functions $K$ is selected to minimize the mean squared error \cite{ramsay2005functional, ferraty2006}. 

Table~\ref{tab:table1} shows the identifiers of the households that have been detected as magnitude outliers (M), shape outliers (S) or evolution outliers (E and $\widetilde{\mbox{E}}$ stand for time series of depths and time series of scaled depths, respectively). 
%A household is considered as M or S if it is a magnitude or shape outlier in more than $95\%$ of the days under analysis. 
Columns 2-5 include the outliers detected using the level data, while columns 6-9 report those found by using the first derivative. 
In what follows, we remark on the key learnings.

\begin{table}[]
\caption{Identifiers of the detected outliers.}
\label{tab:table1}
\centering
\begin{tabular}{@{}lllllllllll@{}}
\toprule
 & & \multicolumn{4}{c}{\textbf{Zero derivative}} & & \multicolumn{4}{c}{\textbf{First derivative}}\\ \midrule 
 & \textbf{Meter id}  & \textbf{M}  & \textbf{S} & \textbf{E} & \textbf{$\widetilde{\mbox{E}}$} &  & \textbf{M}  & \textbf{S} & \textbf{E} & \textbf{$\widetilde{\mbox{E}}$} \\ \midrule

\parbox[t]{1mm}{\multirow{6}{*}{\rotatebox[origin=c]{90}{Voltage}}} & vol$_{5746}$ & & & \checkmark & & & \checkmark & & &  \\

&  vol$_{6139}$ & & & \checkmark & & & \checkmark & & & \\

& vol$_{7901}$ & & & \checkmark & \checkmark & & \checkmark & & & \\

& vol$_{9019}$ & & & \checkmark & & & & & & \\

& vol$_{9922}$ & & & & & & \checkmark & & & \\

& vol$_{7951}$ & & & & & & \checkmark & & & \\ \midrule

\parbox[t]{2mm}{\multirow{3}{*}{\rotatebox[origin=c]{90}{Solar}}} & sol$_{9019}$ & & & & & & & & \checkmark & \\

& sol$_{6139}$ & & & & & & & & \checkmark & \\

& sol$_{3538}$ & & & & & & & & & \checkmark \\

\bottomrule
\vspace{0.05cm}
\end{tabular}

\end{table}

\subsubsection*{Evolution outliers are not detected by other methods}
\begin{figure}
     \centering
     \begin{subfigure}[b]{0.75\textwidth}
         \centering
         \includegraphics[width=\textwidth]{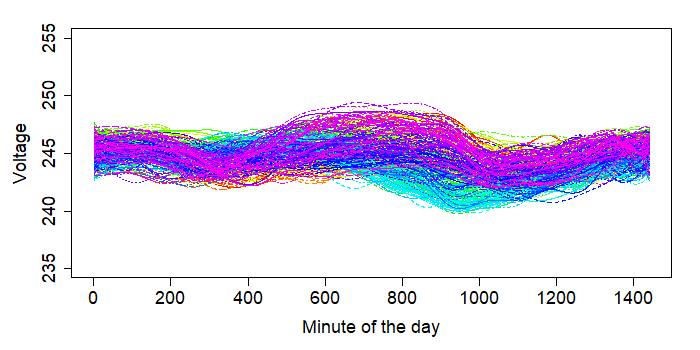}
         \caption{Non-outlier vol$_{2818}$}
         \label{fig:voltage_a}
     \end{subfigure}
     \hfill
     \begin{subfigure}[b]{0.75\textwidth}
         \centering
         \includegraphics[width=\textwidth]{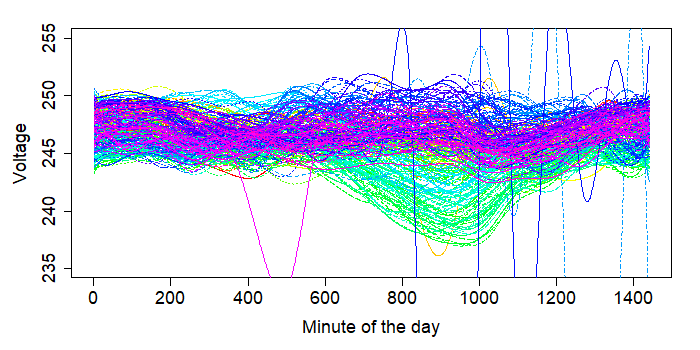}
         \caption{Outlier vol$_{9019}$}
         \label{fig:voltage_b}
     \end{subfigure}
     \hfill
    \begin{subfigure}[b]{0.75\textwidth}
         \centering
         \includegraphics[width=\textwidth]{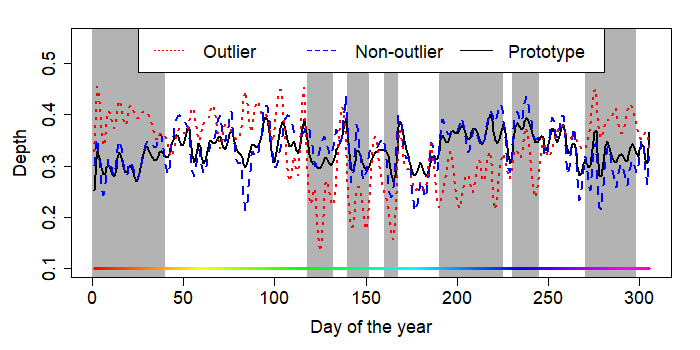}
         \caption{Time series of depths}
         \label{fig:voltage_c}
     \end{subfigure}
        \caption{Voltage circuit: evolution outlier not detected with other method.}
        \label{fig:austin_1}
\end{figure}

As Table~\ref{tab:table1} shows, the methodology proposed in this paper allows us to uncover outliers that are not caught by existing methods for detecting magnitude or shape functional outliers. 
In particular, although meter vol$_{9019}$ is not identified as a magnitude or shape outlier, this household follows an abnormal daily voltage evolution with respect to the group of households and therefore, it is classified as an evolution outlier. 
Subfigure~\ref{fig:voltage_a} shows $305$ daily curves for a non-outlier household (vol$_{2818}$) and Subfigure~\ref{fig:voltage_b} the corresponding daily curves for the detected evolution outlier (vol$_{9019}$).
Each daily curve is colored with a rainbow palette \cite{rainbow2010} associated with the calendar day, that is, similar colors are days which are close in time.

A preliminary visual inspection of Subfigure~\ref{fig:voltage_a} and Subfigure~\ref{fig:voltage_b} reveals the outlying nature of vol$_{9019}$ in comparison with vol$_{2818}$.
They show that voltage daily curves of the same period of time have a different relative magnitude position for the non-outlier and for the outlier. However, one should expect roughly synchronized evolution for two households fed by the same substation branch.
Specifically, the outlier profile has a group of green curves located in low values of voltage, while they are located centrally for the non-outlier. 
Light blue curves are above the majority for the outlier household; and for the non-outlier, they are located below and in the middle of the majority of the curves.

The difference in the evolution is more evident in Subfigure~\ref{fig:voltage_c} where the time series of depths, $\mbox{MBD} (t)$, are represented for the non-outlier and the outlier. Moreover, the prototype, $\mu \mbox{MBD}(t)$, is plotted with a solid line. Here, the outlier (dotted line) moves far away from the prototype, while, in contrast, the non-outlier (dashed line) remains close to it.
%As  shown in the top-two panels, these time series show light blue and green points with lower depths values for the outlier than those for the non-outlier and the prototype.

\begin{figure}
     \centering
     \begin{subfigure}[b]{0.75\textwidth}
         \centering
         \includegraphics[width=\textwidth]{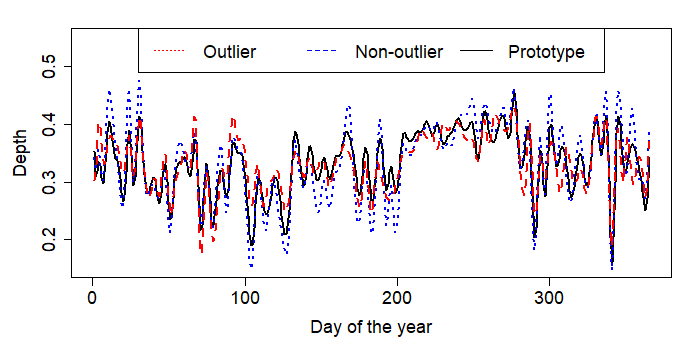}
         \caption{Time series of depths computed on the zero derivative.}
         \label{fig:fig2_a}
     \end{subfigure}
     \hfill
     \begin{subfigure}[b]{0.75\textwidth}
         \centering
         \includegraphics[width=\textwidth]{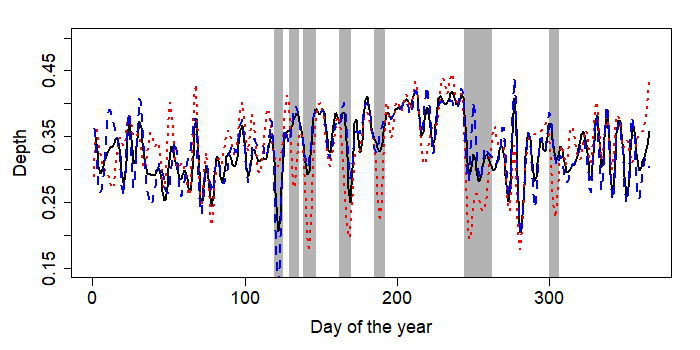}
         \caption{Time series of depths computed on the first derivative.}
         \label{fig:fig2_b}
     \end{subfigure}
     \hfill
	\caption{Photo-voltaic energy generation: Computed depths on the derivatives allow detecting outliers not unmasked by the analysis without derivatives.}
	\label{fig:austin_2}
\end{figure}

\subsubsection*{First derivatives allow detecting those outliers not detected with level data}
Another remark from Table~\ref{tab:table1} is that the use of the first derivatives discloses outliers not unmasked with the functions' values themselves.
This is the case for the circuit voltage of the outlying households vol$_{9922}$ and vol$_{7951}$, which  are two of the just six households that do not have photo-voltaic energy generation. 
The effect of not having solar energy generation on the household circuit voltage is not large enough to be caught with level data, however, the derivatives intensify the shape differences and they are detected as outliers in the magnitude of the derivative. 

Similarly, the first derivatives allow highlighting households with abnormalities in terms of  solar energy generation.
While magnitudes of solar profiles are determined by the amount of power generation installed, the shapes are highly influenced by the panels orientation and tilt. 
In fact, the two households sol$_{6139}$ and sol$_{9019}$, which are detected as evolution outliers (E) in the first derivatives, have their solar panels set to the south, whereas the majority of the households are south-west-oriented\footnote{Panel tilt is not available from the metadata of the Pecan Street data set to have the complete picture of the solar panel setting.}.

For a better understanding of these evolution outliers, Subfigure~\ref{fig:fig2_a} shows the time series of depths of one outlier household (sol$_{6139}$), one non-outlier household (sol$_{4767}$) and the prototype.
The time series of depths for the non-outlier and the outlier are not far  from the prototype, meaning that their daily evolution is fairly similar.  
In contrast, if we consider the first derivatives,  more discrepancies appear. 
To see this, Subfigure~\ref{fig:fig2_b} represents the time depths of the same households and the prototype computed on the first derivatives where the outlier profile is farther from the prototype than the non-outlier (shaded grey regions).

This points to the fact that the analysis of the derivatives might capture the shape differences in the daily generation solar profile due to the differences of panel orientation and tilt.
Therefore, our methodology can be useful, for example, to detect outliers in terms of panel settings when data of a group of meters with a similar panel configuration are available.

\subsubsection*{Scaled depths unmask those outliers which are not detected with regular depths}
%Time series of depths are not able to discern between positive and negative daily trends or peaks and valleys with seasonal data. 
\begin{figure}[]
     \centering
     \begin{subfigure}[b]{0.75\textwidth}
         \centering
         \includegraphics[width=\textwidth]{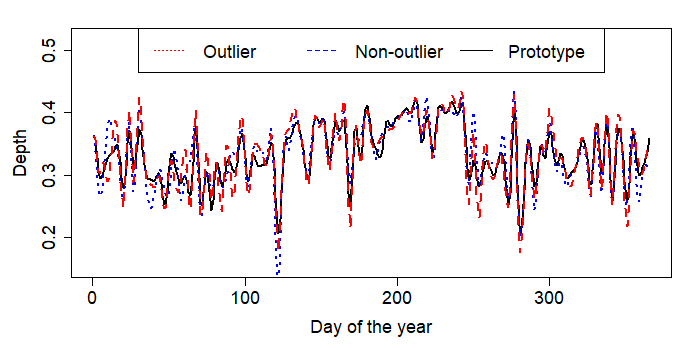}
         \caption{Time series of depths.}
         \label{fig:fig3_nopoints_a}
     \end{subfigure}
     \hfill
     \begin{subfigure}[b]{0.75\textwidth}
         \centering
         \includegraphics[width=\textwidth]{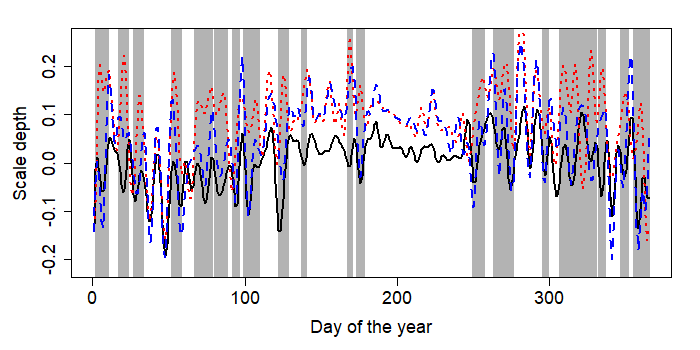}
         \caption{Time series of scaled depths.}
         \label{fig:fig3_nopoints_b}
     \end{subfigure}
     \hfill
	\caption{Photo-voltaic energy generation: Scaled depths detect outliers not detected by classical depths.}
	\label{fig:austin_3}
\end{figure}
%This means that the time depth of a meter with a positive daily trend would behave similarly to the time depth of a meter with a daily negative trend. 
For example, a household with a systematic growth of voltage would provide time depths that are similar to a household whose voltage circuit systematically decreases. In contrast, scaled depths are especially defined to shed light on differences in these variations in trends and seasons.
Table~\ref{tab:table1} shows that the use of scaled depths ($\widetilde{\mbox{E}}$) with photo-voltaic solar energy generation captures the household sol$_{3538}$ as an outlier whereas it is not captured with the other methods, including classical depths. 

Figure~\ref{fig:austin_3} illustrates this particular case. More precisely, Subfigure~\ref{fig:fig3_nopoints_a} shows the regular time depths of the outlier household (sol$_{3538}$) and one non-atypical (sol$_{4767}$) computed on the first derivatives of solar energy generation. 
Both time series of depths are close to the prototype. However, the scaled depths $\widetilde{\mbox{MBD}}$ represented in Subfigure~\ref{fig:fig3_nopoints_b} highlight periods where the atypical is remarkably far from the prototype (shaded grey regions).

Additionally, we see that the $\widetilde{\mbox{MBD}}$ of the outlier is generally above the prototype when the differences with the prototype are large. This means that the solar setting of this household provides a daily profile with larger periods of growth (positive derivatives) than the majority of the households, especially at the start and  end of the year.
In fact, checking the Pecan Street metadata, sol$_{3538}$ is the third largest solar installation facing west and the smallest facing south. This setting produces a double-humped daily profile with a maximum peak of generation that occurs later in the day than that of the majority.

\section{Conclusion}
\label{sec:conclusion}
To fill the absence of methodologies focused on temporal daily dependency for smart meters data \cite{IEEEsun2018, IEEESGwang2019}, we propose an outlier detection method that is able to uncover evolution outliers that remain undetected by current methods. 
The underlying methodology takes advantage of the analysis of multiple grouped meters to extract joint information that affects them equally. 

Furthermore, the pitfalls of not taking into account the time dimension in the task of outlier mining has been shown with actual smart meters data of voltage circuit and solar energy generation.
Using voltage circuit, our proposal captures evolution abnormalities that remain hidden with other methods that are specifically tailored for magnitude and shape outliers. In this context, our approach for grouped meters captures deviations from common dynamics of households fed by the same substation that should have roughly synchronized evolution patterns. 
On the other hand, the temporal evolution plays an important role in the analysis of solar energy generation. Here, only the application of our functional approach to the first derivatives allows highlighting abnormalities due to the differences of panel orientation and tilt. This feature makes our approach an appealing method to monitor solar farms with a given panel configuration or solar tracker systems.

In summary, our outlier detection method proposal, in conjunction with the available methods from the literature, covers a wide and general class of possible atypical phenomena, namely, shape, magnitude, and evolution outliers.
This classification might support in the crucial tasks of monitoring, understanding the sources of the potential abnormality and supporting the decision to intervene.

\bibliographystyle{IEEEtran}
\bibliography{references}


% \end{IEEEbiography}


% % insert where needed to balance the two columns on the last page with
% %\newpage



%\vfill


\end{document}